\documentclass[twocolumn,twoside,slac_two]{revtex4}
\usepackage{graphicx,amsmath}
\usepackage{fancyhdr}
\begin{document}

\title{Polarization in quarkonium production}


\author{J. S. Russ}
\affiliation{Carnegie Mellon University, Pittsburgh, PA 15213, USA}
\author{on behalf of the CDF Collaboration}
%\affiliation{FNAL, Batavia, IL 60510, USA}

\begin{abstract}
Production mechanisms for quarkonium states in hadronic collisions remain difficult to understand.  The decay angular distributions of J/$\psi$ or $\Upsilon(nS)$ states into $\mu^+ \mu^-$ final states are sensitive to the matrix elements in the production process and provide a unique tool to evaluate different models.
This talk will focus on new results for the spin alignment of $\Upsilon(nS)$ states produced in $p\overline{p}$ collisions at $\sqrt{s}$ = 1.96 TeV using the CDF II detector at the Fermilab Tevatron.  The data sample corresponds to
an integrated luminosity of 6.7 fb$^{ -1 }$. The angular distributions are analyzed as functions of the
transverse momentum of the dimuon final state in both the Collins-Soper and the $s$-channel helicity
frames using a unique data-driven background determination method. Consistency of the analysis is checked by comparing frame-invariant quantities derived
from parametrizations of the angular distributions measured in each choice of reference frame. This
analysis is the first to quantify the complete three-dimensional angular distribution of $\Upsilon(1S), \Upsilon(2S)$ and $\Upsilon(3S)$  decays.  The decays are nearly isotropic in all frames, even when produced with large
transverse momentum.

\end{abstract}

\maketitle

\thispagestyle{fancy}


\section{INTRODUCTION} % Section title should be in all capitals.

The physical process by which a heavy quark-antiquark pair is created in a hadronic collision and then emerges bound as a colorless hadronic state has been the subject of intense discussion within the framework of QCD for many years.  The change of color configurations by soft gluon emission was first described in the Color Singlet Model (CSM) by Baier and Ruckl~\cite{csm} in 1983.  Subsequent
cross section measurements at CDF I~\cite{cdf1} in 1994 showed that the predictions underestimated the data by factors of 10-50.  An innovative effective field theory approach (NRQCD) by Bodwin, Braaten, and Lepage~\cite{nrqcd} recast the problem in terms of a perturbative treatment at the quark level with long-range process-independent parameters to describe the hadronization.  The free parameters in the problem were set from the differential cross section data but result in a
definite statement about the contributing matrix elements in the process.  Therefore, the polarization state of the resulting vector meson is determined from the theory.  The prediction is that at large $p_T$ the quarkonium state should be transversely polarized.\\

The prediction refers to quarkonium states that are directly produced in the hard collision.  Experimentally, one measures quarkonium states that are {\it prompt}, i.e., that come from the primary production vertex.  Therefore, the 
experimental data include not only directly-produced quarkonium states but also those that come from the decay of higher-lying $Q\overline{Q}$ excitations, either S-wave or P-wave, a process known as {\it feed-down decays}.  These extra events will almost certainly have different polarizations than the directly produced states.  This presents a fundamental difficulty in interpreting the results from all experiments thus far.  Some theoretical calculations attempt to estimate the feed-down depolarization effects, but there are large uncertainties in these estimates.\\

Recent phenomenological approaches have taken different viewpoints than the
NRQCD approach to describe the quarkonium production process.  In the $k_T$ factorization approach of Baranov and Zotov~\cite{kt}, the color-singlet process dominates, and the heavy quark pair is isolated from the rest of the interaction by a minimum momentum transfer from each parent hadron.  They predict a rather featureless $p_T$ dependence that should be {\it longitudinal} for direct production, as will be seen later.  In contrast, the higher-order CSM approach of Artoisenet, {\it et al.}~\cite{art} considers multiple-gluon interactions and finds that the higher-order processes contribute coherently to the production amplitude and so build up the cross section to a point that generally agrees with current quarkonium observations.  This is a dramatic shift and has reinvigorated interest in this model.\\

In all three cases there are unique predictions for the polarization expected in direct production of quarkonium.  This makes these new Tevatron results on $\Upsilon(nS)$ polarization and the expected new LHC results on quarkonium polarization especially important.\\

\section{PREVIOUS QUARKONIUM POLARIZATION STUDIES}

\subsection{Charmonium Polarization Results}

\begin{figure}[h]
\includegraphics[width=65mm]{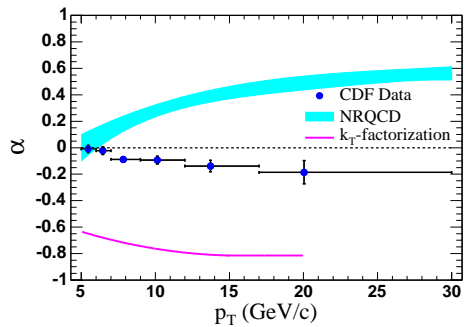}
\caption{Prompt J/$\psi$ Polarization Parameter $\alpha$ in the $s$-channel helicity frame as a function of the dimuon $p_T$.  Theory curves for NRQCD and $k_T$ factorization are included for comparison.  The NRQCD band includes estimates of feed-down depolarization effects.\label{jpsipol}}
\end{figure}

Following the success of NRQCD predictions for the J/$\psi$ and $\psi$(2S) differential cross section results, using the freedom available in matching the long- distance parameters of the model, CDF set out to improve the measurements of
the prompt polarization for these charmonium states using the larger event samples
of Tevatron Run II and the improved measurement accuracy of the CDF II detector.
The prompt polarization for both  J/$\psi$ and $\psi$(2S) does not increase
as $p_T$ becomes significantly larger than the mass of the state, i.e., 
in the perturbative regime.  The results for the J/$\psi$ are shown in Fig.~\ref{jpsipol}.  The measured polarization is small at all $p_T$ values, in
contrast with NRQCD expectations.  The $k_T$ prediction for direct polarization
has a large magnitude, but the shape matches experiment.  Depolarizing effects from feed-down might reduce the discrepancy in size.  More work is needed in this area.  These results are the state of the art at this moment.  New results are expected from LHC experiments soon.\\

\subsection{Bottomonium Polarization Results}

The charm quark mass is close to $\lambda_{QCD}$, and there is some speculation that disagreements with NRQCD predictions might imply that the charmonium system is too light for NRQCD assumptions to be valid.  That is certainly not the case with the bottomonium system, which is the focus of the rest of this report.
Another issue whose importance has been emphasized in recent years is the choice of quantization axis along which to report polarization.  In fixed target experiments the Collins-Soper (CS) axis~\cite{cs}, the bisector of the angle between the
colliding hadrons in the rest frame of the dimuon system, was the axis of choice.  In hadron colliders a natural axis was the $s$-channel helicity axis (SH), defined
as the direction of the Lorentz boost that transforms the dimuon from the lab system to its rest frame.  In elementary quantum mechanics we learn that a fully-polarized electron will have no spin alignment along an axis perpendicular to
its quantization axis.  Braaten {\it et al.}~\cite{bra} emphasized the need to analyze experiments in several different Lorentz frames to ensure that a
small polarization was not an accident of frame choice.  More recently Faccioli {\it et al.}~\cite{facc} have emphasized the importance of measuring the
three independent parameters that describe the parity-conserving decay of a vector meson into a pair of spin-1/2 fermions in the vector meson rest frame:
\begin{eqnarray*}
\frac{dN}{d\Omega} & \propto & \frac{1}{3 + \lambda_{\theta}}[ 1 + \lambda_{\theta} cos^2 \theta + \lambda_{\phi} sin^2 \theta \ cos 2\phi  \\
 & & \mbox{} + \lambda_{\theta \phi} sin 2\theta \ cos \phi]
\end{eqnarray*}

The new results reported here will follow the admonitions from these papers and constitute the first hadron collider results that measure all three polarization parameters in each of two Lorentz frames:  CS and SH.
In addition Ref.~\cite{facc} pointed out a frame-independent quantity that serves both as a systematic constraint on the measurements in two or more Lorentz frames and also, should it be small, indicates that there is no significant polarization in any Lorentz frame.  We shall also discuss that quantity for the new CDF results. \\

\begin{figure}[h]
\includegraphics[angle=-90,width=65mm]{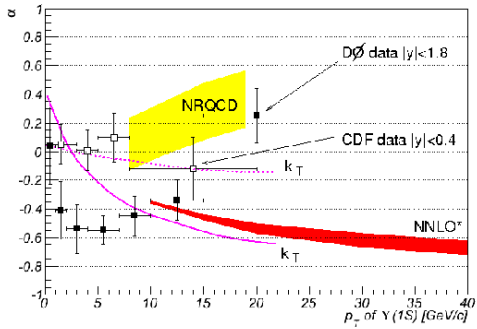}
\caption{$\Upsilon(1S)$ polarization parameters as a function of $p_T$ prior to the new CDF measurement.  The disagreement between CDF Run I and D0 Run II is dramatic and is still not understood.  The two theoretical curves for $k_T$ represent extreme choices for the feed-down effect.  The CSM curve does not include feed-down. \label{ypol1}}
\end{figure}

Just as in the charmonium system, all three production models (NRQCD, $k_T$, and
higher-order CSM) contain enough free parameters to match the $\Upsilon(1S)$ cross section data well over the whole measured $p_T$ range.  The polarization measurement once again stands as the essential test of validity for the model results.   Prior to this new CDF Run II measurement, there were results from CDF Run I~\cite{cdf2} and D0 Run II~\cite{d0}, both in the SH frame.  The two experiments do not agree with each other, and neither experiment matches theory, as can be seen in Fig.~\ref{ypol1}.  The experimental disagreement is not unique in polarization measurements.  These are differential measurements, and good control of acceptance and efficiency is crucial to obtaining stable results.  This experimental fact of life makes using the frame-independent test an important validation of new results.

\begin{figure*}[t]
\centering
\includegraphics*[width=135mm]{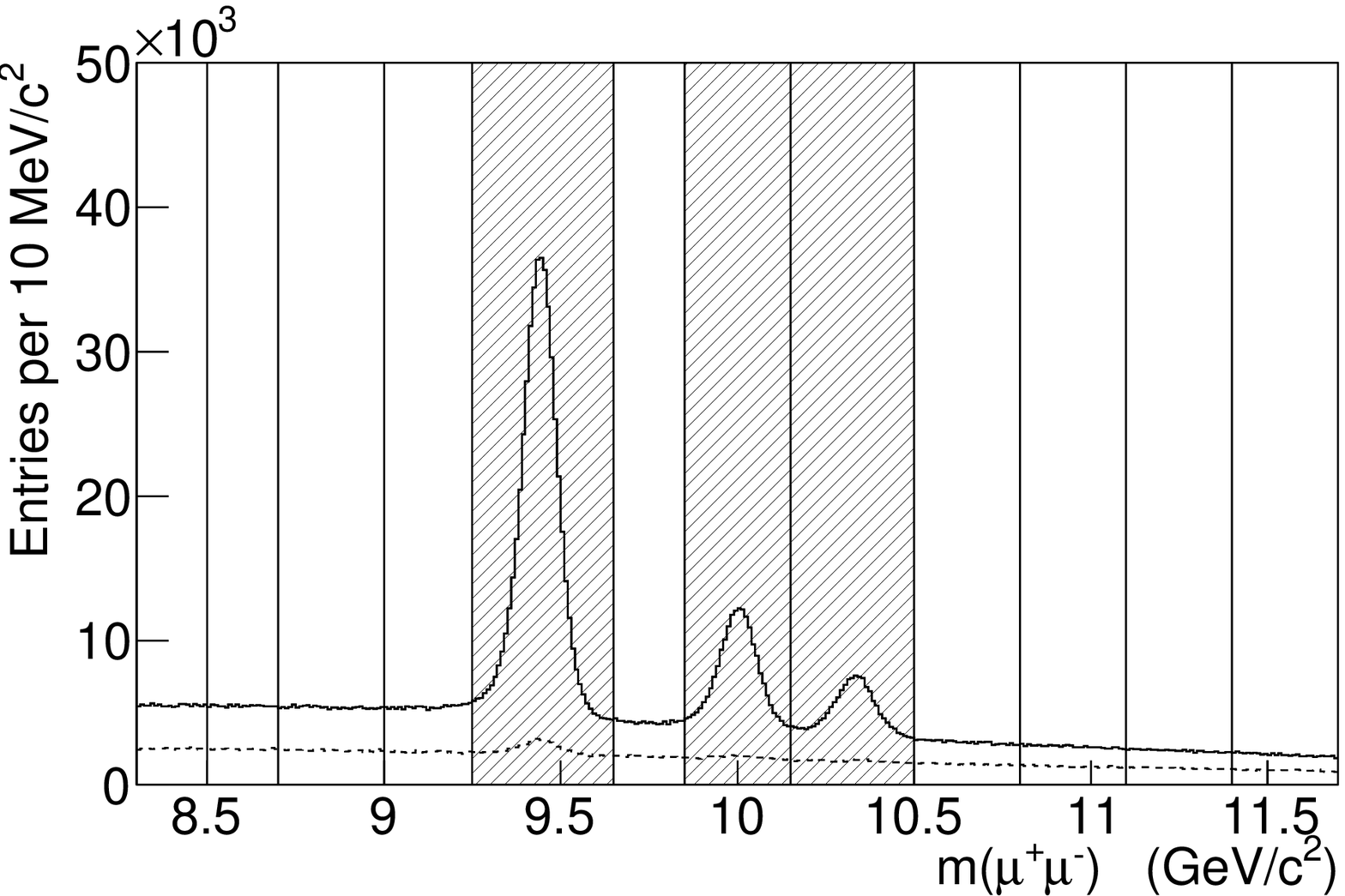}
\caption{CDF dimuon mass spectrum integrated over $p_T$ with $\mid y(\mu \mu) \mid < 0.6$.  Note that the three $\Upsilon(nS)$ peaks are well separated. The different vertical bands denote dimuon mass slices used to study signal and background.\label{cdfmass}}
\end{figure*}

\begin{figure}[h] 
\includegraphics[angle=-90,width=65mm]{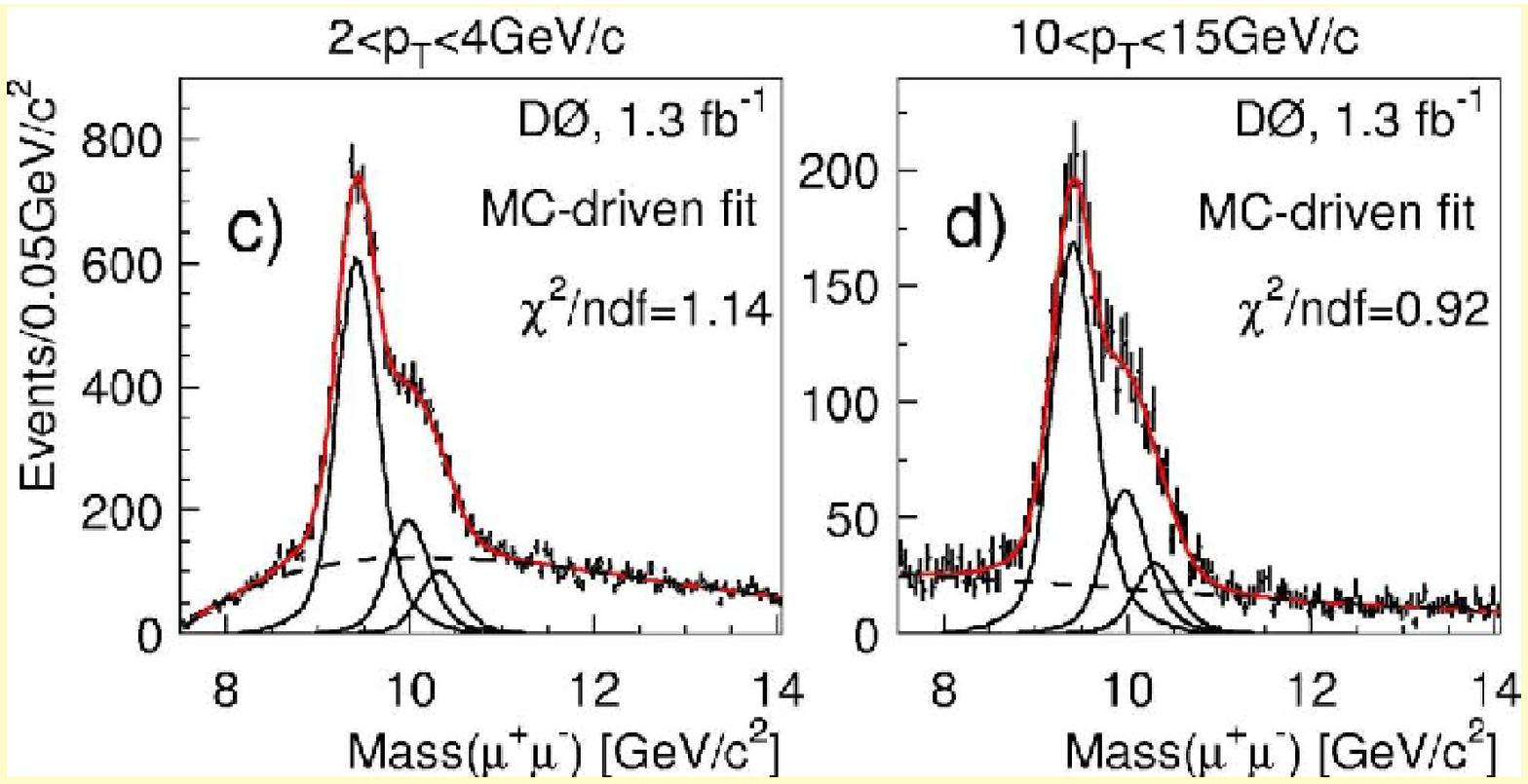}
\caption{D0 dimuon mass spectrum for low and mid-range $p_T$ bins integrated over the rapidity range $\mid y(\mu \mu) \mid < 1.8$.  The three $\Upsilon(nS)$ peaks have $p_T$-dependent overlap.  \label{d0mass}}
\end{figure}

One major difficulty with the $\Upsilon(nS)$ measurements is handling the dimuon background, which is much larger than in the J/$\psi$ case.  The CDF and D0 background
problems are very different, as can be seen in Figs.~\ref{cdfmass} and Fig.~\ref{d0mass}.

\begin{figure}[h]
\includegraphics*[width=65mm]{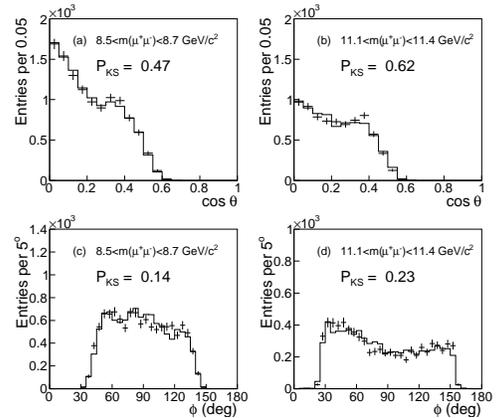}
\caption{Normalized distribution of the CS helicity polarization angle $\theta$ and azimuth $\phi$ for dimuons from the prompt dimuon sample (solid histogram) and the detached dimuon sample (points) for low mass and high mass regions that are pure background.
The agreement in shape validates the claim that the angular characteristics of the combinatoric background are independent of the apparent vertex displacement.
\label{both}}
\end{figure}

While it is clear from the preceding two pictures that the background fraction under any $\Upsilon(nS)$ peak is smaller for CDF than for D0, the importance of that fact really comes from the CDF observation that the combinatoric background has an
angular variation that changes dramatically with dimuon mass and $p_T$.  At the Tevatron the dominant dimuon background comes from heavy flavor decays.  CDF defines a displaced dimuon sample by selecting dimuon-triggered events in which at least one muon has a miss-distance with respect to the primary vertex that exceeds 150 $\mu$m.   This selection defines a sample dominated by combinatoric
dimuons, with a tiny admixture of mis-measured $\Upsilon(nS)$ events.  We divide the dimuon mass region into twelve intervals as shown in Fig.~\ref{cdfmass}: three
regions containing $\Upsilon(nS)$ signals plus background and nine pure background regions.  We make the fundamental assumption that for the combinatoric dimuons the angular distribution is independent of the apparent decay length, $i.e.$, the uncorrelated dimuons can appear to have a vertex at a distance that is uncorrelated with their angular characteristics in their rest frame.  If this hypothesis is true, then the angular characteristics of dimuons in the pure background mass slices will be the same for the prompt and displaced samples.  We have tested this hypothesis in all nine background bins and obtained excellent agreement.  We show as examples the normalized distributions in cos \ $\theta_{CS}$ and $\phi_{CS}$ for two such bins, one at low mass, one at high mass, in Fig.~\ref{both}.  We make these comparisons for all three parameters in both CS and SH frames and see comparably good matches. This agreement shows that we can extract the angular character of
the combinatoric background under each of the three prompt $\Upsilon(nS)$ signal regions by
using the measured displaced sample in that same mass region.  No sideband extrapolation is needed.

\begin{figure}
\includegraphics*[angle=-90,width=65mm]{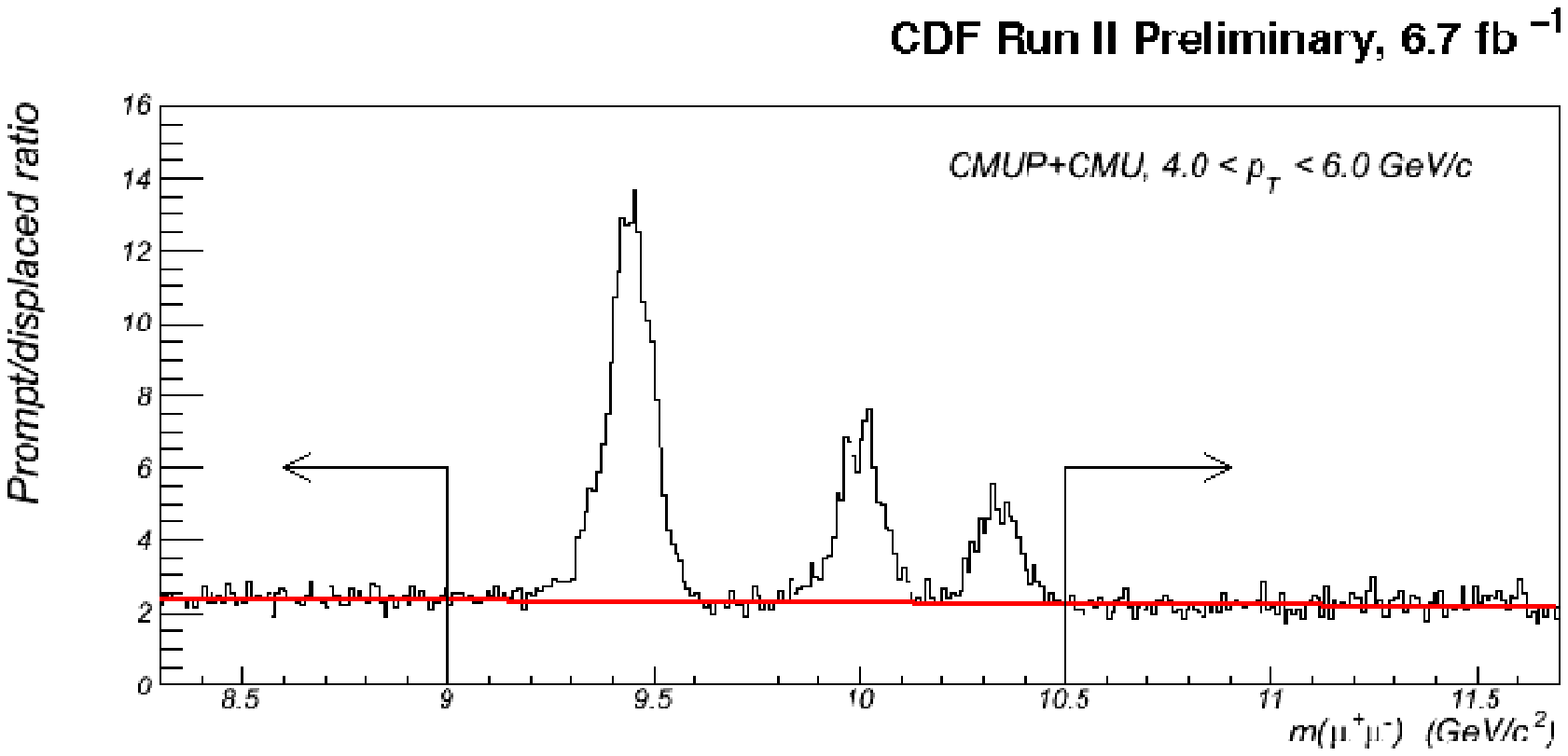}
\caption{Ratio of dimuon mass distribution for prompt and displaced dimuon samples.  The ratio is essentially flat in the background regions.  We use this ratio to scale the displaced vertex background angular distribution in the three $\Upsilon(nS)$ signal bands.  No fit to the signal mass function is needed. \label{ratio}}
\end{figure}

To establish the normalization for the displaced sample, we plot the ratio of the two dimuon samples as a function of dimuon mass, shown in Fig.~\ref{ratio}.  The ratio is essentially
constant in the nine background-only samples.  We fit the ratio to a linear function of dimuon mass and use that fit to scale the displaced sample angular distribution to predict the background prompt angular distribution in the three signal bands.  

 We fit the background angular distributions to the same functional form as the vector meson signal but do not limit the three coefficients for the background to the range expected for a real vector meson.  We make a simultaneous fit to the displaced angular distribution and the prompt angular distribution in each of the twelve dimuon mass ranges.  The prompt angular distribution contains a signal term (if appropriate) and an extrapolated background term, using the ratio as shown above.  All fits give good $\chi^2$ results in both CS and SH frames.  We also included a $cos^4 \theta$ background term.  This term cannot contribute for a real vector meson decay and it plays no role in the polarization analysis.\\

\begin{figure*}[t]
\centering
\includegraphics[angle=-90,width=150mm]{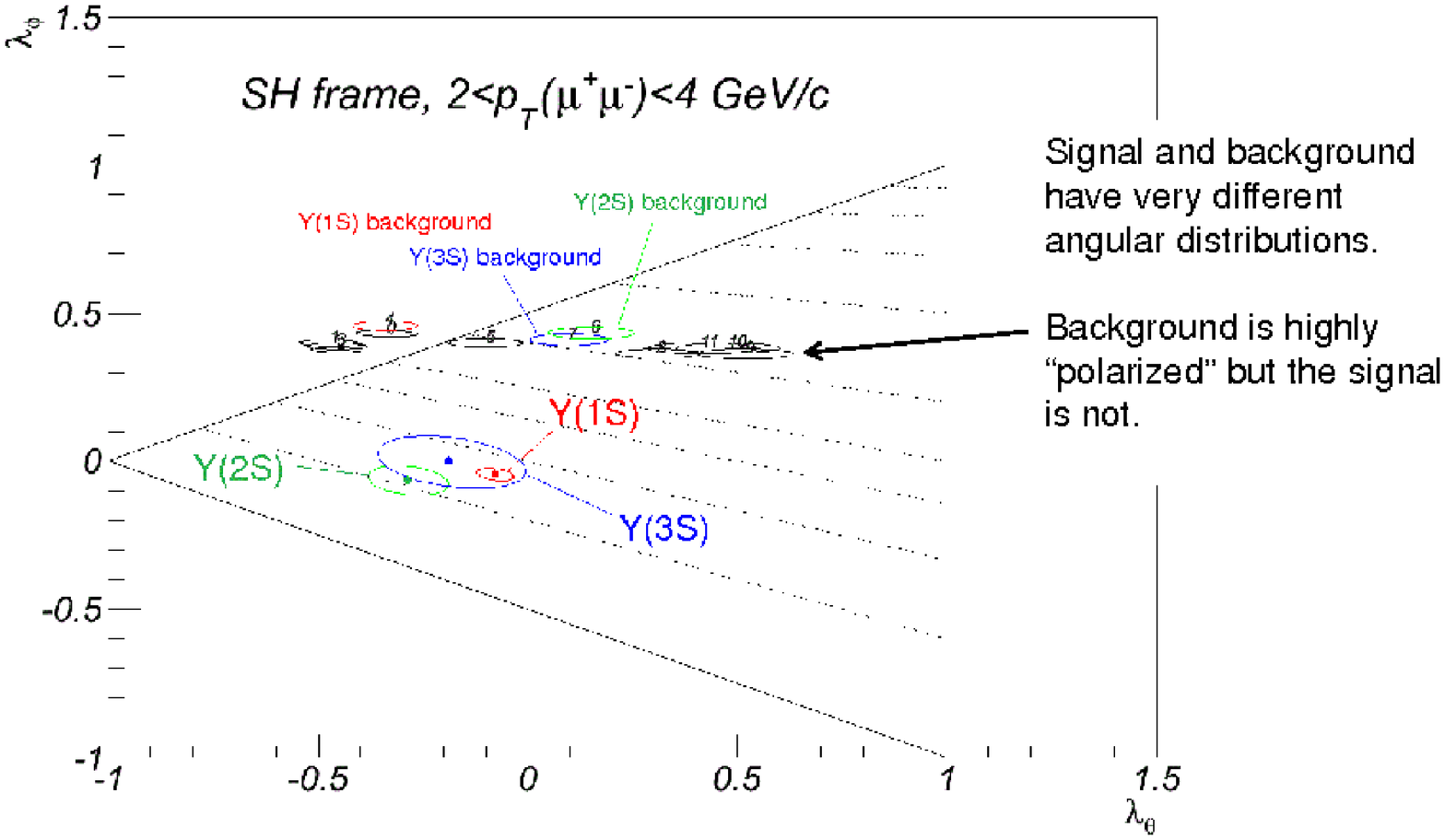}
\caption{Angular fit parameters $\lambda_{\theta}$ vs $\lambda_{\phi}$ for three signal and twelve background prompt dimuon mass regions for 2 GeV/c $< p_T(\mu \mu) <$ 4 GeV/c.  The signal fits cluster near zero for both parameters, while
the background fits tend to have sizeable $\phi$ variation and a mass-dependent
range of $\theta$ variation in the $s$-channel helicity frame shown here. \label{polfit}}
\end{figure*}

An example of the polarization fits is shown in Fig.~\ref{polfit} for the lowest $p_T$ bin.  The twelve dimuon background regions are each represented by a point in the two-dimensional parameter space in $\lambda_{\theta},\lambda_{\phi}$, along with the three signal fits.  Error ellipses for 1$\sigma$ and 2$\sigma$ uncertainties are shown for each point.  As can be seen from the plot, the background polarization parameter $\lambda_{\theta}$ changes non-uniformly with dimuon mass in this $p_T$ region.  Thus, having an accurate representation of background angular behavior from the data itself is essential to extracting the correct signal polarization.  This capability is unique to this analysis.\\

To establish systematic consistency for the measurement, one can use the
frame-independent quantity\\ $\tilde{\lambda}$ = $\frac{\lambda_{\theta} + 3 \lambda_{\phi}}{1 - \lambda_{\phi}}$ defined in Ref.\cite{facc} to compare results from
the CS and SH frames.  Because the same data are used in each frame, one has 
to evaluate matching likelihoods by using a toy Monte Carlo procedure.  In all
$p_T$ bins the results from the two frames agree well, based on the Monte Carlo expectations.

\section{RESULTS}

\begin{figure*}[b]
\centering
\includegraphics[angle=-90,width=150mm]{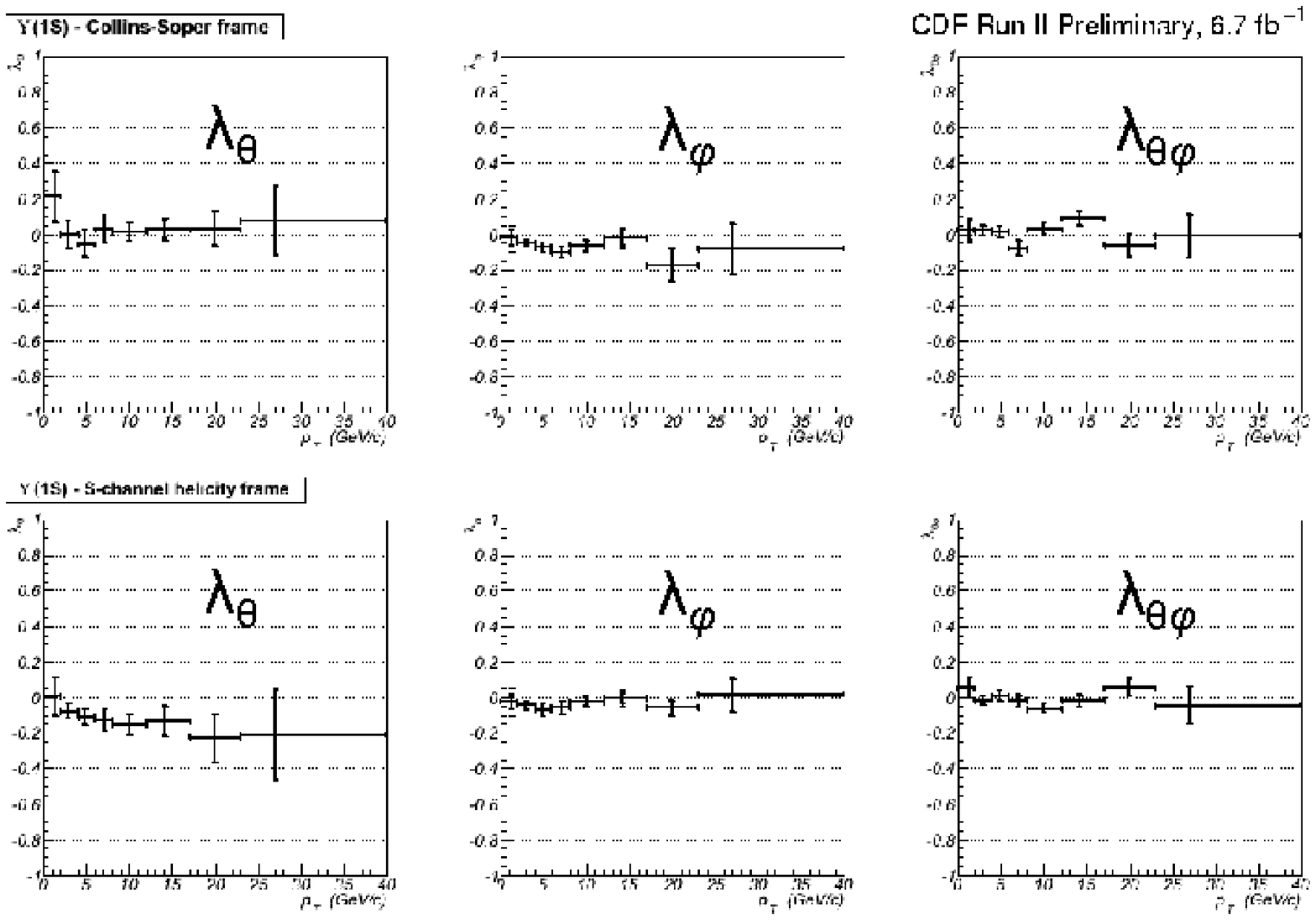}
\caption{Fit results for the three polarization parameters $\lambda_{\theta}, \lambda_{\phi}$ and $\lambda_{\theta \phi}$ for the $\Upsilon(1S)$ signal as a function of $p_T$.  In neither the CS or SH frames is there any significant polarization effect.  Systematic uncertainties are shown as extensions of the statistical error bars on each point. \label{y1pol}}
\end{figure*}

The complete results from this analysis, including tables of all fitted parameters in the CS and SH frames along with the frame-independent parameters are published in the supplemental material to Ref.~\cite{cdf3}.  We give illustrative examples here, starting with those for $\Upsilon(1S)$ prompt decays, shown in Fig.~\ref{y1pol}.

The $\Upsilon(1S)$ polarization parameters are all small and show little variation with $p_T$ in either the CS or SH frames.  The values of the frame independent parameters are also
small and $p_T$-independent, confirming that there is no Lorentz frame in which one would see large polarization for this state.\\

\begin{figure*}[t]
\centering
\includegraphics[angle=-90,width=150mm]{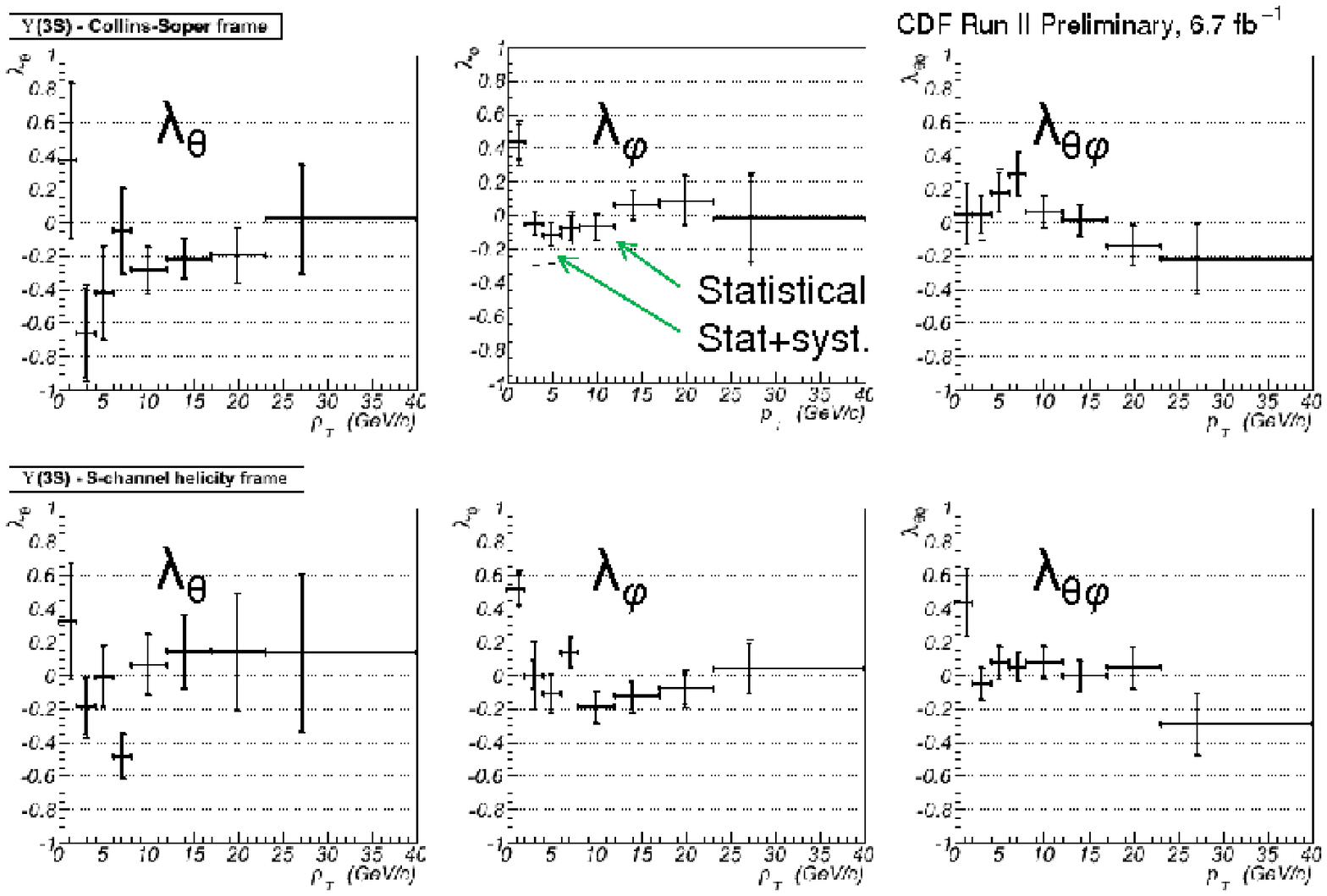}
\caption{Fit results for the three polarization parameters $\lambda_{\theta}, \lambda_{\phi}$ and $\lambda_{\theta \phi}$ for the $\Upsilon(3S)$ signal as a function of $p_T$.  In neither the CS or SH frames is there any significant polarization effect.  Systematic uncertainties are shown as extensions of the statistical error bars on each point. \label{y3pol}}
\end{figure*}

To avoid having to interpret depolarizing effects of feed-down to $\Upsilon(1S)$, we can look at the $\Upsilon(3S)$
parameters.  There is much less feed-down to this state because of smaller
branching fractions from the $\chi_{3b}$ states and because the higher-lying $\Upsilon(nS)$ states have very tiny branching fractions to $\Upsilon(3S)$.
This is the first measurement of polarization parameters for hadroproduction of the $\Upsilon(3S)$ state.  The sample size is smaller than that for the $\Upsilon(1S)$, but
the results shown in Fig.~\ref{y3pol} give no indication of significant polarization at high $p_T$ in either the CS or SH frames.

\begin{figure*}[b]
\centering
\includegraphics*[angle=-90,width=135mm]{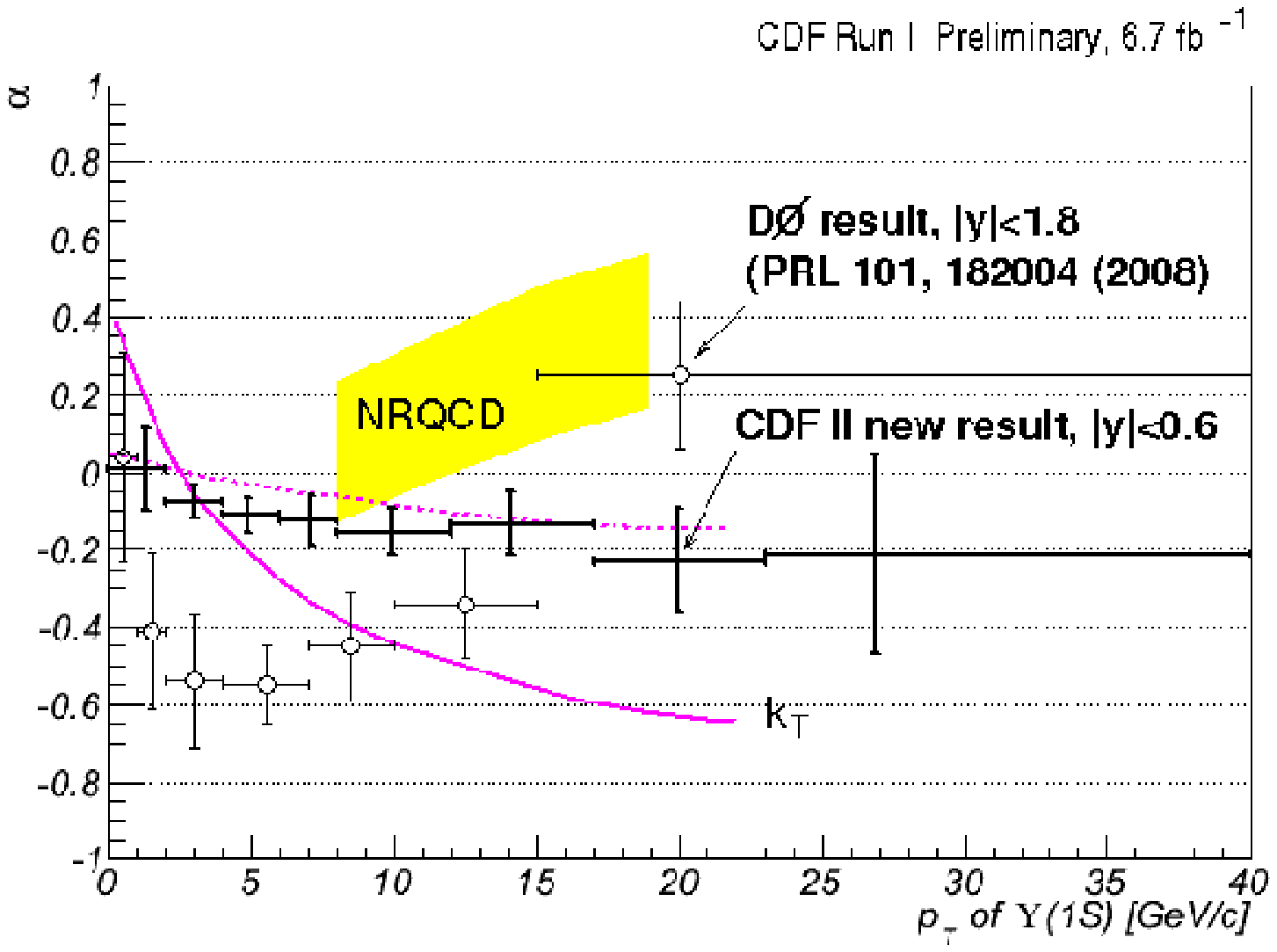}
\caption{$\Upsilon(1S)$ polarization parameters as a function of $p_T$ for the new CDF measurement.  The disagreement with D0 Run II persists.  The theoretical curves for the $k_T$ model is given for two possible feed-down situations, as discussed in Ref.~\cite{kt}. \label{polcomp}}
\end{figure*}

The new CDF results for the $\Upsilon(1S)$ agree well with the CDF Run I measurement and disagree strongly with the D0 measurements, as shown in Fig.~\ref{polcomp}.  The situation for the polarization parameter $\lambda_{\theta}$, also known as $\alpha$, for the SH frame is summarized in Fig.~\ref{polcomp}.  There is no indication of a trend toward NRQCD behavior, illustrated by the yellow band.  Shown on the plot are two curves for the $k_T$ model including feed-down from Ref~\cite{kt}.  The solid line assumes
spin coherence between Y states, while the dashed line shows the case when the feed-down $\Upsilon(1S)$ is depolarized.  The data seem to favor depolarization.  A direct measurement of
$\Upsilon(1S)$ polarization from $\chi_{b1}$ parents is clearly quite important for
understanding the situation.  Not shown here but included in Ref.~\cite{art} is a polarization band for the higher-order CSM that has the same general shape as the data and is consistent with these results.  The width of the uncertainty band can be reduced by having more experimental information about feed-down contributions.\\

\section{CONCLUSIONS}

These new CDF results represent the first hadron collider determination of the complete $\Upsilon(1S), \Upsilon(2S)$ and $\Upsilon(3S)$ polarization parameter set in both Collins-Soper and 
$s$-channel helicity frames for promptly-produced bottomonium decays to $\mu^+ \mu^-$.  The results show that all polarization parameters are small and have little variation with dimuon $p_T$.  The trend of the data tends to support the
$k_T$ model and the higher-order CSM model but not NRQCD expectations.  We await the upcoming analyses from LHC to see the trends at higher energies and higher $p_T$ in pp collisions.

\bigskip % extra skip inserted
\begin{acknowledgments}

We thank the Fermilab staff and the technical staffs of the participating institutions for their vital contributions. This work was supported by the U.S. Department of Energy and National Science Foundation; the Italian Istituto Nazionale di Fisica Nucleare; the Ministry of Education, Culture, Sports, Science and Technology of Japan; the Natural Sciences and Engineering Research Council of Canada; the National Science Council of the Republic of China; the Swiss National Science Foundation; the A.P. Sloan Foundation; the Bundesministerium f\"ur Bildung und Forschung, Germany; the Korean World Class University Program, the National Research Foundation of Korea; the Science and Technology Facilities Council and the Royal Society, UK; the Russian Foundation for Basic Research; the Ministerio de Ciencia e Innovaci\'{o}n, and Programa Consolider-Ingenio 2010, Spain; the Slovak R\&D Agency; the Academy of Finland; and the Australian Research Council (ARC). 

\end{acknowledgments}

\bigskip % extra skip inserted

\begin{thebibliography}{99} % Use for 10-99 references

\bibitem{csm}
Baier, R. and R. Ruckl, ``Hadronic Collisions: A Quarkonium Factory", 
Z. \ Phys. \ C19, 251, 1983.
%      doi            = "10.1007/BF01572254",
%      year           = "1983",
%      reportNumber   = "BI-TP 83/02",
%      SLACcitation   = "%%CITATION = ZEPYA,C19,251;%%",


\bibitem{cdf1}
F. Abe, {\it et al.},
 ``J/$\psi$ and $\psi(2S)$ Production in $p\overline{p}$ Collisions at $\sqrt{s}$ = 1.8 TeV", Phys. Rev. Lett. 79, 572 (1997).


\bibitem{nrqcd}
Bodwin, G. T., E. Braaten and G. P. Lepage, 
  ``Rigorous QCD analysis of inclusive annihilation and production of heavy quarkonium", Phys. Rev. D51, 1125-1171, 1995.


\bibitem{kt}
Baranov, S.P. and N.P. Zotov, ``Upsilonium polarization as a touchstone in understanding the proton dynamics in QCD", JETP Lett. 86, 435-438, 2007. 
%arXiv:hep-ph 0707.0253.

\bibitem{art}
      Artoisenet, P., J.M. Campbell, J.P. Lansberg, F. Maltoni and F. Tramontano, ``$\Upsilon$ Production at Fermilab Tevatron and LHC Energies", Phys. Rev. Lett. 101, 152001, 2008. 

\bibitem{cs}
Collins, J. and D. Soper, ``Angular Distribution of Dileptons in High-Energy Hadron Collisions", Phys. Rev. D16, 2219, 1979.

\bibitem{bra}
Braaten, E., D. Kang, J. Lee and C. Yu, ``Optimal spin quantization axes for quarkonium with large transverse momentum", Phys. Rev. D79, 054013, 2009. 
%arXiv: hep-ph 0812.3727.

\bibitem{facc}

Faccioli, P., C. Lourenco and J. Seixas, ``Rotation-invariant Relations in Vector Meson Decays into fermion pairs", Phys. Rev. Lett. 105, 061601, 2010. 
%arXiv: hep-ph 1005.2601.

\bibitem{cdf2}

 Acosta, D. {\it et al.}, ``$\Upsilon$ production and polarization in $p\bar{p}$ collisions at $\sqrt{s}=$ 1.8-TeV", Phys. Rev.Lett.88,161802,2002.

\bibitem{d0}
Abazov, V.M., {\it et al.}, ``Measurement of the polarization of the $\Upsilon_{1S}$ and $\Upsilon_{2S}$ states in $p \bar{p}$ collisions at $\sqrt{s}$ = 1.96-TeV", Phys. Rev. Lett. 101, 182004, 2008. 
%arXiv: hep-ex 0804.2799.

\bibitem{cdf3}
Aaltonen, T., {\it et al.}, ``Measurements of Angular Distributions of Muons From
$\Upsilon$ Meson Decays in $p\bar{p}$ Collisions at $\sqrt{s}=1.96$ TeV",
Phys. Rev. Lett. 108, 151802, 2012. 
%arXiv: hep-ex 1112.1591.


%\bibitem{accelconf-ref}
%http://www.cern.ch/accelconf

%\bibitem{}

\end{thebibliography}
%\begin{thebibliography}{9}   % Use for  1-9  references

\end{document}